\documentclass[manuscript,screen,nonacm]{acmart}
\usepackage{array}
\usepackage{longtable}
\usepackage{multirow}
\AtBeginDocument{%
  }

\setcopyright{none}
\begin{document}

\title[Beyond Average Gains]{Beyond Average Gains: Probing the Equity--Vulnerability Tension in Structured Visual AI for Learning}

\author{Mingyuan Li}
\affiliation{%
  \institution{University of Pennsylvania}
  \city{Philadelphia}
  \state{PA}
  \country{USA}
}
\email{myuli@engineering.upenn.edu}

\author{Duan Wang}
\affiliation{%
  \institution{University of California, San Diego}
  \city{La Jolla}
  \state{CA}
  \country{USA}
}
\email{duw003@ucsd.edu}

\author{Erick Purwanto}
\affiliation{%
  \institution{Xi'an Jiaotong-Liverpool University}
  \city{Suzhou}
  \country{China}
}
\email{Erick.Purwanto@xjtlu.edu.cn}

\author{Yihong Wang}
\authornote{Corresponding author.}
\affiliation{%
  \institution{Xi'an Jiaotong-Liverpool University}
  \city{Suzhou}
  \country{China}
}
\email{Yihong.Wang@xjtlu.edu.cn}

\renewcommand{\shortauthors}{Li et al.}

\begin{abstract}
Open-ended conversational AI requires learners to formulate questions, organize learning paths, and track changing problem states. Structured visual interfaces may ease these demands, but higher average ratings do not establish that support is experienced consistently across learner backgrounds and devices. We conducted a randomized study of graph-algorithm learning, comparing VisualCodeChat (VCC), which combines predefined learning stages, conversational explanations, and dynamic algorithm visualizations, with a standard text-based chatbot. We analyzed 282 post-task questionnaires from university students. In unadjusted comparisons, VCC received significantly higher usability and perceived effectiveness ratings. We detected no significant between-condition differences in efficiency, satisfaction, or post-use emotion. Exploratory within-condition analyses identified differences associated with prior chatbot experience: infrequent users reported lower adoption and learning-management ratings in the standard AI condition, while uncomfortable prior chatbot experiences were associated with lower interaction and layout ratings in the VCC condition. These findings suggest that improving the average interaction experience is only part of making AI-supported learning inclusive. Future evaluations should examine whether learners with different prior experiences can obtain guidance, recover from difficulties, and demonstrate understanding independently across devices.
\end{abstract}



\maketitle

\section{Introduction}

AI-supported learning systems increasingly provide students with on-demand explanations, guidance, and feedback. The accessibility of this support can vary with the interaction demands learners face. This concern is especially relevant in graph-algorithm learning, where students need to reason about abstract structures while tracking changes in algorithmic states. Algorithm visualizations can make these processes more observable and support conceptual understanding \cite{hundhausen_2002_MetaStudyAlgorithm,naps_2002_ExploringRole}, while conversational AI allows students to request explanations and guidance in natural language \cite{kazemitabaar_2024_CodeAidEvaluating}. Open-ended interaction, however, requires learners to formulate useful questions, determine what to ask next, and connect textual explanations with changing algorithmic states. These demands may make the quality of AI-supported learning partly dependent on learners' prior familiarity with conversational systems. This raises the question of whether a system rated favorably on average provides comparably accessible support to learners with different levels of prior chatbot use.

Structured visual AI offers one possible response to this problem. Rather than requiring learners to organize the interaction entirely through free-form prompting, a structured interface can divide learning into predefined stages, provide stepwise conversational guidance, and synchronize explanations with visual representations of algorithm execution. Such designs build on long-standing evidence that well-integrated verbal and visual representations can support learning \cite{mayer_2003_NineWays,naps_2002_ExploringRole}. By shifting part of the work of organizing a learning path and tracking algorithmic state from the learner to the interface, structured visual AI may reduce experience-related barriers to engaging with AI support. We use the term \textit{interactional equity} narrowly to refer to whether learners with different levels of prior chatbot use report comparably usable and supportive interactions. This framing concerns perceived access to the interaction rather than broader educational equity or equality in learning outcomes.

Reducing an initial interaction barrier, however, does not imply that support remains equally effective across all learner backgrounds and use conditions. Prior affective experiences can shape how learners approach and interpret technology-mediated learning \cite{dmello_2012_DynamicsAffective,pekrun_2006_ControlvalueTheory}, while the conditions under which a technology is accessed can constrain how its interface is experienced \cite{warschauer_2010_NewTechnology}. A structured interface may make a learning path clearer, yet its fixed organization and visual presentation may also be less accommodating when learners need additional clarification, revisit earlier material, or interact through different devices. These possibilities suggest a distinction between lowering the barrier to entering an AI-supported learning interaction and sustaining useful support once learners are engaged. We therefore treat the equity--vulnerability relationship as a potential evaluative tension: design features that reduce one source of interactional difficulty may still leave other experience- or context-dependent limitations unresolved. We examine whether higher average ratings coexist with differences associated with learners' prior experiences and conditions of use.

Prior research provides substantial evidence for the instructional value of algorithm visualization, the role of affect in learning, and the importance of technological context \cite{hundhausen_2002_MetaStudyAlgorithm,naps_2002_ExploringRole,dmello_2012_DynamicsAffective,pekrun_2006_ControlvalueTheory,warschauer_2010_NewTechnology}. These perspectives, however, motivate examining not only whether an AI learning interface receives higher average evaluations, but also how those evaluations are distributed across learners with different prior experiences and devices. We therefore ask two questions. First, compared with a conventional text-based AI interface, how does structured visual AI differ in perceived usability, effectiveness, efficiency, satisfaction, and post-use emotion? Second, within each interface condition, how do learners' evaluations vary with prior chatbot-use frequency, prior affective experience with chatbots, and device type?

To investigate these questions, we conducted a controlled study in a graph-algorithm learning setting and analyzed 282 post-task questionnaires. Participants were randomly assigned to either a VisualCodeChat (VCC) condition or a standard AI condition and completed the same sequence of three graph-algorithm learning tasks. Both conditions used the same Large Language Model (LLM) and instructional sequence, while VCC combined predefined learning stages, conversational explanations, and dynamic algorithm visualizations; the standard AI condition provided a conventional text-based conversational interface. After completing the tasks, participants reported their perceived usability, effectiveness, efficiency, satisfaction, and emotional experience. We additionally examined prior chatbot-use frequency, prior affective experience with chatbots, and device type as learner and context variables. These measures capture self-reported interaction experience and perceived learning support rather than independently assessed learning gains.

The results show a selective, rather than uniform, advantage for structured visual AI. Participants using VCC reported significantly higher usability and perceived effectiveness than those using standard AI, whereas we detected no significant between-condition differences in efficiency, satisfaction, or post-use emotion. Exploratory within-condition analyses further revealed variation that was obscured by these average comparisons. In the standard AI condition, infrequent chatbot users reported lower User Adoption and Learning Management evaluations, whereas no corresponding significant frequency differences were reported within VCC. Within VCC, participants with previously uncomfortable chatbot experiences gave lower ratings on several interaction-related dimensions, and laptop users reported higher Interface ratings than tablet or smartphone users. These subgroup patterns do not establish that VCC eliminates experience-related disparities or that it is more strongly affected by prior emotional experience; rather, they show that average improvements can coexist with background- and context-associated differences in perceived support.

This study makes three contributions. Empirically, it characterizes where structured visual AI received higher ratings, particularly in usability and perceived effectiveness. Conceptually, it distinguishes lowering experience-related barriers to entering an AI-supported interaction from sustaining comparable support across different prior experiences and use conditions. This distinction frames the equity--vulnerability tension as a question for future evaluations. Finally, the findings motivate design directions such as recoverable interaction paths, mechanisms for requesting localized assistance, cross-device evaluation, and independent assessment of learning outcomes.

\section{Related Work}

\subsection{Educational AI and the Pursuit of Equity}
Educational AI can provide scaffolding tailored to different learners. Scaffolding builds on Vygotsky's concept of the zone of proximal development \cite{vygotsky_1978_MindSociety} and describes how guidance helps learners complete tasks they could not yet perform independently \cite{wood_1976_RoleTutoring}. Intelligent tutoring systems and chatbots are common forms of AI-supported scaffolding \cite{holmes_2022_StateArt}. These technologies can provide personalized feedback during self-directed learning. For example, AI-powered language learning tools can offer customizable and low-cost opportunities for practice \cite{rebolledofontdelavall_2023_ExploringBenefits}. \citet{son_2023_ArtificialIntelligence} discuss text- and speech-based AI tools for language learning. Students have reported that using ChatGPT supports their foreign language learning, particularly in writing, grammar, and vocabulary \cite{karatas_2024_IncorporatingAI}. AI-assisted instruction is also widely used in computing education. For example, using ChatGPT to support programming instruction can enhance students' computational thinking skills and learning motivation \cite{yilmaz_2023_EffectGenerative}. Tools such as CodeAid use an LLM to guide students in solving programming problems themselves \cite{kazemitabaar_2024_CodeAidEvaluating}.

Equity in AI-scaffolded learning is frequently framed as reducing barriers related to prior experience, skill level, or instructional support \cite{melo-lopez_2025_ImpactArtificial}. AI tools can help learners access material beyond their current language proficiency, address gaps in foundational knowledge, and navigate disability-related barriers \cite{kohnke_2025_ArtificialIntelligence}. Research on AI and educational equity often emphasizes access and fairness across demographic groups, including gender, cultural background, and socioeconomic status \cite{roshanaei_2023_HarnessingAI}. Learner states such as motivation, cognitive load, and emotional engagement may also shape usability and learning outcomes \cite{dmello_2012_DynamicsAffective}. Examining equity in practice therefore requires considering how learners experience AI support during use.

\subsection{Structured Visual Design in Learning}
Visual scaffolding in AI-supported learning environments has emerged as a powerful strategy to enhance learning outcomes by reducing cognitive load and fostering conceptual understanding. Research demonstrates that visual aids, such as interactive diagrams and simulations, help learners process complex information more effectively by offloading cognitive demands \cite{mayer_2003_NineWays,rau_2016_ConditionsEffectiveness}. Structured learning pathways externalize abstract processes to support novice learners. For instance, by breaking down complex algorithms into step-by-step sequences, learners can obtain knowledge more systematically \cite{hundhausen_2002_MetaStudyAlgorithm}. For example, step-based animations assist students’ algorithm learning by showing the algorithm decision through next-step visualization \cite{grivokostopoulou_2017_EducationalSystem}. However, most studies on visual scaffolding and structured pathways assume stable and ideal learning conditions, where learners are emotionally engaged, attentive, and working in optimal contexts. They often overlook the impact of variable learner states, such as emotional disengagement, distraction, or fatigue, which can significantly affect learning outcomes. Previous research show that negative emotions in digital environments have a detrimental effect on learning outcomes \cite{wortha_2019_MultipleNegative}. Additionally, learning effectiveness can be compromised by suboptimal learning environments such as using outdated devices or facing unreliable internet access \cite{reich_2017_GoodIntentions}. While structured visual design promotes inclusivity through cognitive accessibility, it may carry hidden costs when emotional and situational variability are introduced.

\subsection{Emotion, Usability, and the Fragility of Interaction}
Learners experience dynamic emotional fluctuations during deep learning activities \cite{dmello_2012_DynamicsAffective}. Emotions are of primary educational importance, emotional comfort plays a critical role in shaping cognitive load, attention, and learning persistence in educational contexts \cite{pekrun_2006_ControlvalueTheory}. Studies show that learners in emotionally supportive environments demonstrate better learning performance as positive emotions facilitate deeper cognitive processing and motivation \cite{kim_2013_EmotionsMotivation}. Conversely, negative emotions such as anxiety, hopelessness, anger and boredom can lead learners to lose interest in learning. In the field of affective computing, studies reveal that negative emotions like frustration, confusion, or disengagement significantly degrade perceived learning system usability \cite{arguel_2017_OutDetecting}. For example, studies on digital learning systems suggest that confusion can trigger frustration and, over time, boredom \cite{arguel_2017_OutDetecting}, ultimately reducing users’ trust in and perceived effectiveness of the system. However, most educational AI systems do not explicitly account for the emotional dimension of interaction. While some affect-aware systems incorporate affective computing techniques, such as facial expression analysis or physiological sensors to detect emotions \cite{malekzadeh_2015_ReviewEmotion}, the majority rely on standardized instructional designs that assume stable emotional states. These systems typically calculate an evaluation score for an educational AI system based on system performance, learner outcomes, and prior knowledge \cite{mousavinasab_2018_IntelligentTutoring}, without considering how emotional states like frustration or disengagement might influence these metrics. Structured tutoring systems are vulnerable because they may fail to adapt learners’ learning state when they deviate from the preset pathways. For example, the educational system proposed in \cite{grivokostopoulou_2017_EducationalSystem} presents a fixed instructional sequence, guiding learners through each algorithmic step with interactive animations. However, a misunderstanding at any point can hinder comprehension of later steps, and the system lacks functionality to review or repeat earlier stages. As a result, users may be unable to complete subsequent tasks such as practice exercises, which leads to breakdowns in interaction and a decline in overall user experience.

\subsection{Contextual Robustness and the Hidden Assumptions of Design}

The design and evaluation of contemporary learning technologies are often grounded in idealized assumptions about learners' contexts. These systems frequently presume a ``digitally affluent'' environment—characterized by reliable hardware, high-speed internet access, and ample screen space \cite{warschauer_2010_NewTechnology}. Such assumptions, however, obscure the socio-technical entanglement of educational technology, wherein technology use is embedded within learners’ messy and dynamic everyday lives \cite{orlikowski_2008_SociomaterialityChallenging}. Consequently, when interfaces optimized for large screens and reliant on complex visual metaphors are transferred to mobile devices, it is not merely the layout that fails, but the underlying cognitive interaction models \cite{giannakos_2013_UsingWebcasts}. More alarmingly, validity assessments in computer education research have long favored sterile laboratory settings, creating a methodological contextual blindness. This evaluative paradigm fails to capture authentic user interactions under non-ideal conditions, such as emotional distress, cognitive fatigue, or environmental distractions, which are characteristic of real-world learning \cite{strijbos_2010_DevelopingRole}. As a result, systems that perform well in sanitized environments and appear ``equitable'' in theory may rapidly degrade in practical deployment, inadvertently exacerbating rather than mitigating educational inequality \cite{scott_2015_SAGEHandbook}. These serve as cautions that discourse on educational equity must move beyond mere access to address the ``robustness-in-practice'' of technology in diverse, real-world situations.

\subsection{Toward Emotionally Resilient Educational AI}
In response to these research gaps, this study shifts the evaluative focus from universal design to contextual adaptability, investigating the consistency of performance of structured, visualized AI-based educational tools across diverse learner backgrounds and affective states. We posit that while many AI-driven educational tools aim to enhance equity by providing highly structured scaffolding to lower cognitive entry barriers, this strategy conceals an under-examined risk. However, this approach may embed an overlooked risk, which we term the equity–vulnerability tension: the pursuit of initial usability and accessibility (equity) may compromise a system’s resilience and flexibility under complex or stressful conditions (vulnerability). When learners encounter frustration or anxiety, their cognitive resources are depleted \cite{pekrun_2006_ControlvalueTheory}, and the rigid scaffolds intended to assist may become cognitive constraints, limiting adaptive responses, and undermining the development of self-regulation \cite{koedinger_2007_ExploringAssistance}. The contribution of this study lies in introducing a more dynamic and ecologically valid evaluation lens for equity-focused design. We argue that genuine equity should not be confined to static support structures but should encompass systems with adaptive stability—the capacity to sensitively respond to fluctuations in learners’ cognitive and emotional states, and to maintain effective support in variable real-world conditions. Thus, we advocate a conceptual shift in educational design from static usability to dynamic resilience.

\section{Methodology}

\subsection{Research Procedure}

The experiment was conducted as part of the “Advanced Object-Oriented Programming” course at Xi’an Jiaotong-Liverpool University (XJTLU), involving 320 third-year students with prior programming experience. This background enabled them to engage with relatively complex algorithmic concepts. The experimental content focused on three graph algorithm topics: Graph Connectivity, Identifying Connected Components in an Undirected Graph, and Detecting Cycles in a Directed Graph. These topics represent canonical graph problems of comparable difficulty, encompassing both directed and undirected graphs. They require computational thinking and the comprehension of abstract algorithmic principles, making them suitable for evaluating the instructional interventions under investigation. Ethical approval for this study was obtained from XJTLU. 

VCC, the conversational agent component of VisualCodeMOOC \cite{li_2025_VisualCodeMOOCCourse}, served as the instructional tool in the structured visual AI condition. It guided students through the graph-algorithm learning tasks using a predefined sequence of conceptual explanations, step-by-step walkthroughs with dynamic visualizations, code explanations, and practice questions. The visualizations illustrated algorithmic steps alongside the chatbot’s explanations. During the interaction, students could confirm their understanding, ask follow-up questions, and receive clarification and feedback on their answers. To isolate the influence of the structured visual interaction design, both conditions were powered by the same LLM model through the same ChatGPT 3.5 API and model configuration. The two conditions were matched in learning duration, instructional topics and their order, learning content and tasks, and the amount and timing of instructor guidance. In both conditions, participants received a complete demonstration for the first topic, partial guidance for the second topic, and conducted self-directed learning for the third topic. The only intended difference was the interaction and presentation format: VCC organized the learning process into predefined stages and coordinated chatbot explanations with dynamic visualizations, whereas the standard AI condition delivered the corresponding learning support through a conventional text-based conversational interface. Consequently, differences between the conditions were interpreted as being associated with the interaction and presentation format rather than the underlying language model, instructional content, exposure time, or instructor support.

Participants were randomly assigned to one of two groups: the VCC group or the standard AI group. To ensure experimental control and minimize cross-group influence, participants completed the study in one of eight lab sessions conducted at different times. At the beginning of each session, the researcher provided standardized instructions outlining the learning tasks and experimental procedure. Each session lasted approximately 50 minutes and followed the same sequence of learning topics: (1) graph connectivity, (2) identifying connected components, and (3) detecting cycles. After completing all learning tasks, participants completed a post-study questionnaire.

For the standard AI group, instructor-assisted guidance included a brief explanation of algorithmic concepts and step-by-step guidance on how to use the AI tool for learning. In the first topic, a complete demonstration of how to use the AI tool to learn the algorithm was provided. For the second topic, students received partial guidance before independently exploring the remaining learning of algorithm detailed steps. In the third topic, students engaged in self-directed learning using the AI tool. Similarly, for the VCC group, the first topic involved a full demonstration of the system's interface and interactive learning process. In the second topic, instructors guided students through the first two stages—understanding the algorithmic concept and its basic steps—after which students explored the remaining algorithm steps independently. The third topic was conducted as a fully self-directed learning task. A total of 320 questionnaire responses were collected. Following data cleaning, 38 responses were excluded due to indicators of potentially invalid responding, such as uniform response patterns across questionnaire items or unusually short completion times. The final analytical sample comprised 282 responses. Among the 282 participants with complete demographic data, 83 were women and 203 were men. Their ages ranged from 18 to 26 years ($M = 21.09$, $SD = 0.91$).

\subsection{Measurement Instruments and Adaptation}
This study employed a randomized, between-subjects, post-test-only design to compare two AI-supported learning conditions: VCC and a standard text-based AI chatbot. The assigned AI condition constituted the independent variable. The primary dependent variables were participants’ post-use ratings of usability, perceived effectiveness, efficiency, satisfaction, and emotion, measured using 50 items on a five-point Likert scale. Demographic and learner-background information was collected to characterize the sample, while prior chatbot-use frequency, prior affective experience with chatbots, and device type were examined as subgroup variables. Prior affective experience referred to participants’ previous perceptions of chatbot use, whereas the Emotion scale assessed their affective responses to the assigned system following the learning session. All five outcome dimensions represented participants’ self-reported perceptions rather than objective learning performance.

This questionnaire we designed is built upon a comprehensive, multi-dimensional framework to systematically evaluate the holistic performance of an AI chatbot in a specific educational context. Its core structure draws upon the classic usability standard, ISO 9241-11 \cite{tullis_2013_MeasuringUser}, structuring the assessment around the fundamental dimensions of Effectiveness, Efficiency, and Satisfaction. Building on this foundation, the questionnaire deeply integrates the core tenets of the Technology Acceptance Model (TAM) \cite{davis_1989_PerceivedUsefulness} by measuring Perceived Usefulness (reflected in effectiveness) and Perceived Ease of Use (reflected in usability and efficiency) to predict the user's intention for continued use. Furthermore, to achieve a more holistic evaluation, the design transcends traditional functional considerations by incorporating a modern User Experience (UX) perspective. Drawing from models like the User Experience Questionnaire (UEQ) \cite{laugwitz_2008_ConstructionEvaluation}, it establishes a dedicated Emotion dimension. This dimension investigates the tool's impact on learner attributes such as interest, motivation, and confidence, while also assessing its ability to mitigate frustration.

Overall, this questionnaire contains a total of 63 items. It first collects data on participant demographics, their background, including programming experience and past AI chatbot experience, their learning style, their usage context (study location and device), and their motivations for using AI tools. The instrument's core consists of five sections: Usability, Effectiveness, Efficiency, Satisfaction, and Emotion, each containing 10 questions rated on a 5-point Likert scale (1 = Strongly Disagree; 5 = Strongly Agree). These sections are designed to respectively evaluate the tool's ease of operation, the value of its content for aiding learning, its performance in system response and simplifying the learning process, the user's overall satisfaction and intent for continued use, and its positive affective impact on the learner's motivation and confidence.

\subsection{Validity and Reliability of the Questionnaire}
To assess the construct validity of the adapted questionnaire, an exploratory factor analysis (EFA) was first conducted via SPSS. This began with a principal components analysis (PCA), in line with Kaiser’s recommendation \cite{kaiser_1960_ApplicationElectronic} that retained factors should have eigenvalues greater than 1.00. In addition, Cattell’s scree plot criterion \cite{cattell_1966_ScreeTest} was applied. Both eigenvalues and scree plots were used to determine the appropriate number of factors to extract. Subsequently, a PCA with varimax rotation was performed to further explore potential factor structures. Based on the guidelines proposed by DeCoster \cite{decoster_1998_OverviewFactor}, the following criteria were used to retain factors: (1) each item should have a loading greater than 0.4; (2) items should load on only one factor; (3) the retained factors and items should be conceptually coherent; and (4) each factor should contain at least three items.

To assess the internal consistency reliability of the extracted factors, Cronbach’s alpha ($\alpha$) coefficients were computed. Following the widely accepted threshold of $0.70$ \cite{nunnally_1994_PsychometricTheory}, this study adopted $0.70$ as the minimum criterion for acceptable reliability. Usability ($\alpha = .842$, 6 items), Effectiveness ($\alpha = .823$, 7 items), Efficiency ($\alpha = .815$, 7 items), Satisfaction ($\alpha = .935$, 8 items), and Emotion ($\alpha = .876$, 5 items) all exceeded the acceptable threshold.

\begin{figure}[htbp]
    \centering
    \includegraphics[width=1\linewidth]{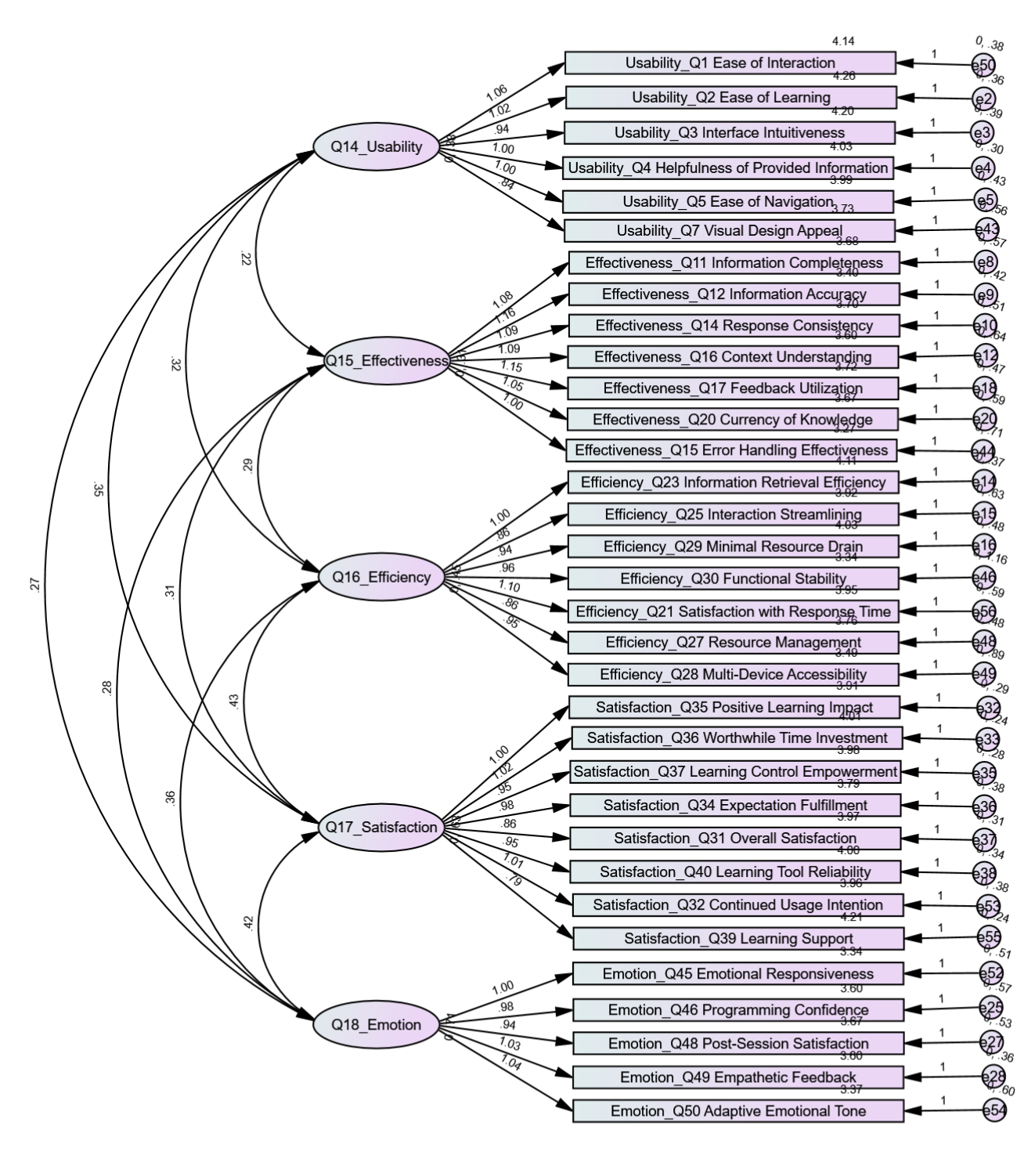}
    \caption{Confirmatory Factor Analysis of the Five User-Experience Scales}
    \Description{A path diagram of five correlated factors, Usability, Effectiveness, Efficiency, Satisfaction, and Emotion, with their 33 questionnaire items and estimated paths.}
    \label{fig:cfa}
\end{figure}

To further examine the factorial validity of the questionnaire constructs, confirmatory factor analysis (CFA) was performed using structural equation modeling (SEM) with AMOS. Model fit was evaluated based on the following commonly used indices: (1) the chi-square to degrees of freedom ratio should satisfy $\chi^2/df < 5$; (2) the Root Mean Square Error of Approximation should satisfy $\mathrm{RMSEA} < 0.08$; (3) the Standardized Root Mean Square Residual should satisfy $\mathrm{SRMR} < 0.08$; and (4) both the Tucker–Lewis Index and the Comparative Fit Index should satisfy $\mathrm{TLI} > 0.90$ and $\mathrm{CFI} > 0.90$, respectively. The results of the CFA indicated an acceptable model fit: $\chi^2 = 942.113$, $df = 485$, $\chi^2/df = 1.943$, $\mathrm{SRMR} = 0.0546$, $\mathrm{RMSEA} = 0.058$, $\mathrm{TLI} = 0.901$, and $\mathrm{CFI} = 0.909$. The corresponding structural equation model is illustrated in Figure \ref{fig:cfa}.

\subsection{Data Analysis} 
Questionnaire data were analyzed using SPSS and AMOS. Following data cleaning, scale scores were calculated from the retained items for the five evaluation dimensions, with higher scores indicating more positive evaluations. Measurement quality was examined through exploratory factor analysis, confirmatory factor analysis, and Cronbach’s alpha. Pearson correlations were used to examine associations among the five dimensions. A one-way multivariate analysis of variance (MANOVA) was conducted to test the overall effect of AI condition on Usability, Effectiveness, Efficiency, Satisfaction, and Emotion. Follow-up independent-samples $t$-tests were used to identify differences in the individual dimensions, with partial $\eta^2$ and Cohen’s $d$ reported as effect sizes where available.

To examine group differences across the chatbot evaluation dimensions, a two-step analytical approach was employed. First, separate MANOVAs were conducted within each AI condition to assess the overall effects of the subgroup variables across multiple dependent variables simultaneously. Subgroup variables that did not yield statistically significant multivariate effects were excluded from further analysis to maintain a focus on meaningful differences. For subgroup variables with significant MANOVA results, follow-up univariate analyses of variance (ANOVAs) were conducted for each relevant dimension. Bonferroni-adjusted post hoc comparisons were then used to determine which groups differed on the dependent variables. When Levene’s test indicated a violation of the homogeneity-of-variance assumption ($p < .05$), the corresponding ANOVA was rerun under adjusted assumptions, and the Games–Howell post hoc test was used to accommodate unequal variances and sample sizes. Cohen’s $d$ was reported for pairwise effect sizes. We used a significance threshold of $\alpha = .05$ and two-sided $t$-tests.

\section{Findings}

We first examined differences between the two AI conditions across the five user-experience scales, followed by correlations among the scales. We then conducted within-condition analyses to examine whether participants' evaluations varied with prior chatbot-use frequency, prior affective experience with chatbots, and device type. All significant subgroup comparisons are summarized in Table~\ref{tab:subgroup-results}.

\subsection{Overall Differences in User Experience between Standard AI and VCC}

\begin{figure}[htbp]
    \centering
    \includegraphics[width=0.82\linewidth]{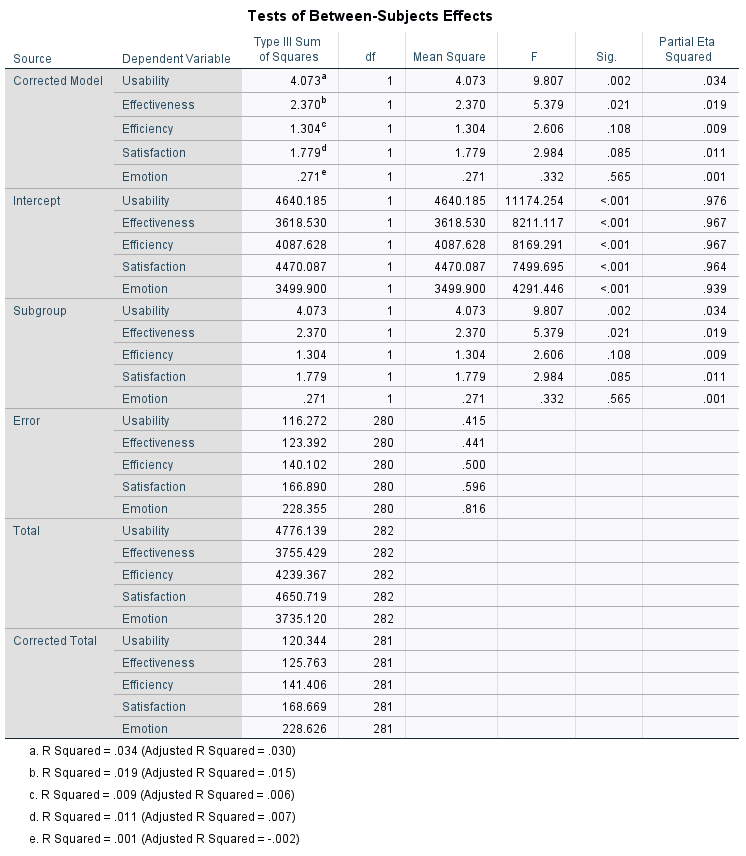}
    \caption{Tests of Between-Subjects Effects for the Five User-Experience Dimensions}
    \Description{An SPSS table reporting Type III sums of squares, degrees of freedom, mean squares, F statistics, significance values, and partial eta squared for Usability, Effectiveness, Efficiency, Satisfaction, and Emotion. The subgroup effect is significant for Usability and Effectiveness but not for the other three dimensions.}
    \label{fig:condition-comparison}
\end{figure}

A one-way MANOVA was first conducted to examine the joint effect of AI condition across the five user-experience dimensions. Figure~\ref{fig:condition-comparison} shows the univariate tests of between-subjects effects. Based on unadjusted $p$ values, the tests indicated differences at the nominal .05 threshold for Usability, $F(1, 280) = 9.81$, $p = .002$, partial $\eta^2 = .034$, and Effectiveness, $F(1, 280) = 5.38$, $p = .021$, partial $\eta^2 = .019$. Participants assigned to VCC reported higher Usability and Effectiveness than those assigned to the standard AI condition.

No statistically significant between-subjects effects were found for Efficiency, $F(1, 280) = 2.61$, $p = .108$, partial $\eta^2 = .009$; Satisfaction, $F(1, 280) = 2.98$, $p = .085$, partial $\eta^2 = .011$; or the post-use Emotion scale, $F(1, 280) = 0.33$, $p = .565$, partial $\eta^2 = .001$. Descriptively, ratings in the standard AI condition had smaller standard deviations than those in the VCC condition across all five scales.

\begin{figure}[htbp]
    \centering
    \includegraphics[width=\linewidth]{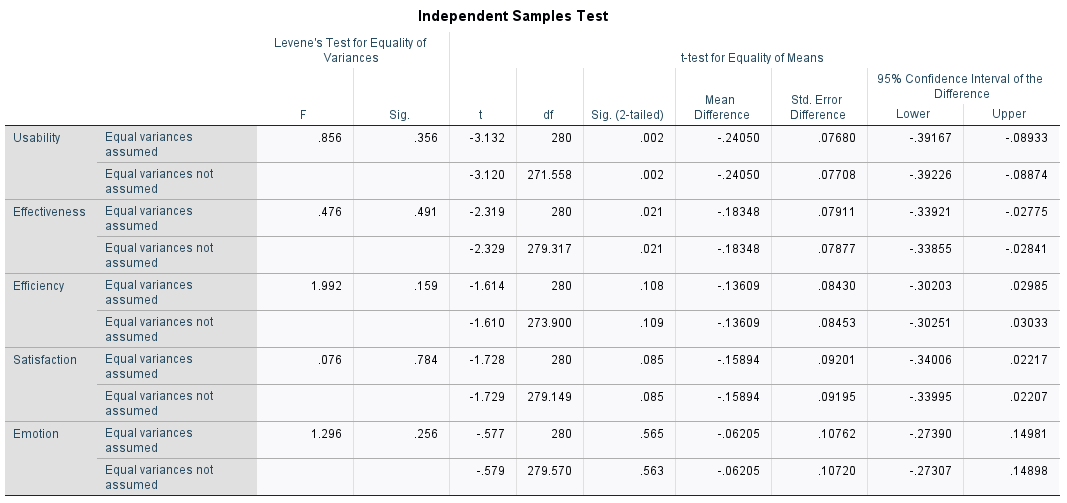}
    \caption{Independent-Samples Tests Comparing Standard AI and VCC}
    \Description{An SPSS independent-samples test table reporting Levene's tests and t-tests for Usability, Effectiveness, Efficiency, Satisfaction, and Emotion. Group differences are significant for Usability and Effectiveness but not for Efficiency, Satisfaction, or Emotion.}
    \label{fig:independent-samples-tests}
\end{figure}

Follow-up independent-samples $t$-tests (Figure~\ref{fig:independent-samples-tests}) reproduced this pattern. Because the mean differences were calculated as standard AI minus VCC, negative values indicate higher ratings for VCC. The difference was significant for Usability, $t(280) = -3.132$, $p = .002$, mean difference $= -0.24050$, $SE = 0.07680$, 95\% CI $[-0.39167, -0.08933]$, and Effectiveness, $t(280) = -2.319$, $p = .021$, mean difference $= -0.18348$, $SE = 0.07911$, 95\% CI $[-0.33921, -0.02775]$. The difference was not statistically significant for Efficiency, $t(280) = -1.614$, $p = .108$, mean difference $= -0.13609$, 95\% CI $[-0.30203, 0.02985]$; Satisfaction, $t(280) = -1.728$, $p = .085$, mean difference $= -0.15894$, 95\% CI $[-0.34006, 0.02217]$; or Emotion, $t(280) = -0.577$, $p = .565$, mean difference $= -0.06205$, 95\% CI $[-0.27390, 0.14981]$. Levene's tests were nonsignificant for all five scales ($p = .159$--$.784$), supporting the equal-variance estimates reported here.

\subsection{Relationships among User-Experience Dimensions}

\begin{figure}[htbp]
    \centering
    \includegraphics[width=0.82\linewidth]{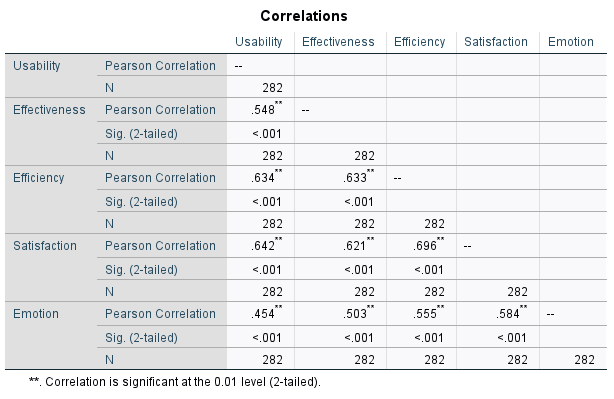}
    \caption{Pearson Correlations among the Five User-Experience Dimensions}
    \Description{A correlation matrix showing statistically significant positive correlations among Usability, Effectiveness, Efficiency, Satisfaction, and Emotion.}
    \label{fig:scale-correlations}
\end{figure}

All five scales were positively correlated, with all pairwise associations reaching statistical significance at $p < .001$ (Figure~\ref{fig:scale-correlations}). Usability was associated with Effectiveness ($r = .548$), Efficiency ($r = .634$), Satisfaction ($r = .642$), and Emotion ($r = .454$). Effectiveness was associated with Efficiency ($r = .633$), Satisfaction ($r = .621$), and Emotion ($r = .503$). Efficiency had the strongest observed association with Satisfaction ($r = .696$) and was also associated with Emotion ($r = .555$). Satisfaction was positively associated with Emotion ($r = .584$).

For the subgroup analyses, separate MANOVAs were conducted within each AI condition for prior chatbot-use frequency, prior affective experience with chatbots, and device type. Significant multivariate effects were followed by univariate ANOVAs and adjusted pairwise comparisons. The omnibus tests, descriptive statistics, assumption checks, and post hoc procedures are summarized in Table~\ref{tab:subgroup-results}; the text below focuses on the significant pairwise comparisons.

\begingroup
\small
\setlength{\tabcolsep}{3pt}
\renewcommand{\arraystretch}{1.15}
\begin{longtable}{|>{\raggedright\arraybackslash}p{0.11\textwidth}|>{\raggedright\arraybackslash}p{0.14\textwidth}|>{\raggedright\arraybackslash}p{0.345\textwidth}|>{\raggedright\arraybackslash}p{0.345\textwidth}|}
\caption{ANOVA Summary by Dimension and AI Condition}\label{tab:subgroup-results}\\
\hline
\textbf{Dimension} & \textbf{Question} & \textbf{Group1 (Standard AI)} & \textbf{Group2 (VCC)} \\
\hline
\multirow[t]{4}{0.11\textwidth}{\textbf{Usability}}
& \textbf{Interaction}
& \textbf{Past Chatbot Feeling}\par Participants who felt neutral reported statistically significantly lower interaction scores than those who felt comfortable.
& \textbf{Past Chatbot Feeling}\par Participants who felt uncomfortable reported statistically significantly lower interaction scores than those who felt neutral. \\
\cline{2-4}
& \textbf{User Adoption}
& \textbf{Past Chatbot Experience}\par Participants who used the chatbot rarely reported statistically significantly lower User Adoption scores than those who used it once in a while, often, and every day.\par \textbf{Past Chatbot Feeling}\par Participants who felt neutral reported statistically significantly lower user adoption scores than those who felt comfortable.
& Null \\
\cline{2-4}
& \textbf{Interface}
& Null
& \textbf{Chatbot Device}\par Participants who used a laptop reported statistically significantly higher interface scores compared to those who used a tablet or smartphone.\par \textbf{Past Chatbot Feeling}\par Participants who felt uncomfortable reported statistically significantly lower interface scores than those who felt comfortable. \\
\cline{2-4}
& \textbf{Layout}
& \textbf{Past Chatbot Feeling}\par Participants who felt neutral reported statistically significantly lower layout scores than those who felt comfortable.
& \textbf{Past Chatbot Feeling}\par Participants who felt uncomfortable reported statistically significantly lower layout scores than those who felt neutral and comfortable. \\
\hline
\multirow[t]{2}{0.11\textwidth}{\textbf{Effectiveness}}
& \textbf{Information Completeness}
& \textbf{Past Chatbot Feeling}\par Participants who felt neutral reported statistically significantly lower information completeness scores than those who felt comfortable.
& \textbf{Past Chatbot Feeling}\par Participants who felt uncomfortable reported statistically significantly lower information completeness scores than those who felt neutral. \\
\cline{2-4}
& \textbf{Response Consistency}
& \textbf{Past Chatbot Feeling}\par Participants who felt neutral reported statistically significantly lower response consistency scores than those who felt comfortable.
& Null \\
\hline
\multirow[t]{2}{0.11\textwidth}{\textbf{Efficiency}}
& \textbf{Response Time}
& \textbf{Past Chatbot Feeling}\par Participants who felt neutral reported statistically significantly lower response time scores than those who felt comfortable.
& \textbf{Past Chatbot Feeling}\par Participants who felt uncomfortable reported statistically significantly lower response time scores than those who felt neutral and comfortable. \\
\cline{2-4}
& \textbf{Information Retrieval}
& \textbf{Past Chatbot Feeling}\par Participants who felt uncomfortable and neutral reported statistically significantly lower information retrieval scores than those who felt comfortable.
& Null \\
\hline
\textbf{Satisfaction}
& \textbf{Learning Management}
& \textbf{Past Chatbot Experience}\par Participants who used the chatbot rarely reported statistically significantly lower Learning Management scores than those who used it often, and every day.\par \textbf{Past Chatbot Feeling}\par Participants who felt neutral reported statistically significantly lower learning management scores than those who felt comfortable.
& \textbf{Past Chatbot Feeling}\par Participants who felt uncomfortable reported statistically significantly lower learning management scores than those who felt comfortable. \\
\hline
\end{longtable}
\endgroup

\subsection{Prior Chatbot-Use Frequency and Evaluations of Adoption and Learning Management}

Prior chatbot-use frequency was associated with two evaluations in the standard AI condition. Participants who rarely used chatbots reported lower User Adoption scores than participants who used them once in a while (mean difference $= -1.091$, $p < .05$, Cohen's $d = 0.7603$), often (mean difference $= -1.117$, $p < .01$, $d = 0.7603$), or every day (mean difference $= -1.486$, $p < .001$, $d = 0.7603$). Participants who rarely used chatbots also reported lower Learning Management scores than those who used them often (mean difference $= -0.942$, $p < .05$, $d = 0.7092$) or every day (mean difference $= -1.382$, $p < .01$, $d = 0.7092$). No significant frequency-related comparison was reported for VCC.

\subsection{Prior Affective Experience and Evaluations across User-Experience Dimensions}

Prior affective experience with chatbots was associated with evaluations of both systems, although the location of the significant pairwise differences varied by condition. In the standard AI condition, significant differences generally occurred between participants who reported neutral and comfortable prior experiences. In the VCC condition, significant differences were concentrated among participants who reported uncomfortable prior experiences.

\subsubsection{Interaction and Interface Evaluations}

For the standard AI condition, participants with neutral prior chatbot experiences reported lower Interaction scores than those with comfortable experiences (mean difference $= -0.514$, $p < .05$, $d = 0.5211$). They also reported lower User Adoption (mean difference $= -0.646$, $p < .01$, $d = 0.6074$) and Layout scores (mean difference $= -0.573$, $p < .05$, $d = 0.4885$) than participants with comfortable prior experiences. No significant affect-related comparison for Interface was reported for this condition.

For VCC, participants with uncomfortable prior chatbot experiences reported lower Interaction scores than those with neutral experiences (mean difference $= -1.099$, $p < .01$, $d = 0.5800$). The same group rated the Interface lower than participants with comfortable experiences (mean difference $= -0.703$, $p < .05$, $d = 0.4304$). They also reported lower Layout scores than participants with neutral experiences (mean difference $= -0.941$, $p < .05$, $d = 0.4843$) or comfortable experiences (mean difference $= -0.827$, $p < .05$, $d = 0.4843$). No significant affect-related comparison for User Adoption was reported for VCC.

\subsubsection{Information Quality and Efficiency Evaluations}

In the standard AI condition, participants with neutral prior chatbot experiences rated Information Completeness lower than those with comfortable experiences (mean difference $= -0.882$, $p < .001$, $d = 0.8189$). They also gave lower ratings for Response Consistency (mean difference $= -0.577$, $p < .05$, $d = 0.5528$) and Response Time (mean difference $= -0.813$, $p < .01$, $d = 0.6385$). For Information Retrieval, participants with uncomfortable (mean difference $= -1.200$, $p < .01$, $d = 1.0263$) or neutral prior experiences (mean difference $= -1.000$, $p < .001$, $d = 1.0263$) reported lower scores than those with comfortable experiences.

For VCC, participants with uncomfortable prior chatbot experiences reported lower Information Completeness than those with neutral experiences (mean difference $= -1.033$, $p < .05$, $d = 0.4487$). They also rated Response Time lower than participants with neutral (mean difference $= -1.013$, $p < .05$, $d = 0.4665$) or comfortable experiences (mean difference $= -0.960$, $p < .05$, $d = 0.4665$). No significant affect-related comparisons were reported for Response Consistency or Information Retrieval in the VCC condition.

\subsubsection{Learning Management Satisfaction}

In the standard AI condition, participants with neutral prior chatbot experiences reported lower Learning Management scores than those with comfortable experiences (mean difference $= -1.017$, $p < .001$, $d = 0.9012$). In the VCC condition, participants with uncomfortable prior experiences reported lower Learning Management scores than those with comfortable experiences (mean difference $= -0.810$, $p < .05$, $d = 0.4669$).

\subsection{Device Type and Interface Evaluations}

Device type was associated with Interface ratings in the VCC condition. Participants who used a laptop rated the interface more highly than those who used a tablet or smartphone (mean difference $= 0.770$, $p < .05$, $d = 0.4304$). No significant device-related comparison was reported for the standard AI condition.

\section{Discussion}

This study compared standard AI and VCC, in terms of usability, perceived effectiveness, efficiency, satisfaction, and post-use affective experience. It also examined how system evaluations varied with prior chatbot-use frequency, prior affective experience with chatbots, and device type. The discussion addresses three questions: what structured scaffolding supports and where its benefits are limited, how these benefits vary across learner backgrounds, and how conditions of use constrain inclusive design.

\subsection{Benefits and Limits of Scaffolding}

VCC received significantly higher ratings for usability and perceived effectiveness than standard AI, whereas no significant overall differences were observed in efficiency, satisfaction, or post-use affective experience. The benefits of structured visual support were therefore concentrated in particular dimensions of experience, rather than constituting a general improvement across all aspects of user experience.

The usability advantage may reflect VCC's structured instructional sequence and closely integrated visualizations. Predictable learning steps can clarify how to proceed, while step-by-step visualizations provide contextual feedback that makes abstract content, such as algorithmic processes, more concrete and observable. This interpretation resonates with Naps et al.'s account of the educational value of algorithm visualization when it actively engages learners \cite{naps_2002_ExploringRole}. In contrast, the more open-ended dialogue and textual explanations of standard AI may require learners to formulate their own queries and track changes in algorithmic states. Sweller's work on instructional structure and cognitive load offers a complementary explanation: insufficiently guided or poorly structured materials can increase extraneous cognitive load \cite{sweller_1994_CognitiveLoad}. VCC's usability advantage may thus reflect the value of shifting some of the work of organizing learning and tracking states to the interface, although this cognitive mechanism was not directly measured.

The advantage in perceived effectiveness can likewise be understood through external representations and mental model construction. Visualization can connect abstract descriptions with the dynamic behavior of computational processes, supporting learners' conceptual understanding of algorithms \cite{hundhausen_2002_MetaStudyAlgorithm}. Segmenting the learning process and externalizing state changes may reduce the working-memory demands of tracking transient information, leaving more cognitive resources available for understanding the underlying logic. This account is consistent with cognitive load theory \cite{chandler_1991_CognitiveLoad}. Moreno's work on interactive multimedia learning also supports the potential contribution of step-by-step engagement to understanding and mental model construction \cite{moreno_2007_InteractiveMultimodal}. Because effectiveness was assessed through learner self-reports, however, the findings indicate greater perceived support for learning, rather than demonstrated gains in knowledge acquisition, transfer, or long-term learning.

Structured scaffolding may also introduce interaction constraints. One possible explanation for the nonsignificant efficiency difference is that a step-by-step sequence provides a clear learning path without necessarily accommodating rapid queries or direct access to specific difficulties. Brusilovsky's work on adaptive hypermedia suggests that guided paths can support novices, whereas experienced users or those with specific goals may place greater value on nonlinear access and control \cite{brusilovsky_2001_AdaptiveHypermedia}. Learning to interpret a visualization also requires time and cognitive effort \cite{sweller_1998_CognitiveArchitecture}. These demands may limit the extent to which usability benefits translate into perceived efficiency, although the present study did not directly test the roles of user control or the effort required to learn the visual representations.

The satisfaction findings further suggest that clarity is only one basis for evaluating an interaction. A constrained learning sequence, unmet interaction expectations, or a reduced sense of control may shape users' overall assessments of a structured system, as suggested by research on user control and satisfaction \cite{obrien_2008_WhatUser}. The familiarity and flexibility of standard AI may also contribute to its experiential value. In Hassenzahl's account, user experience encompasses both instrumental and non-instrumental qualities; improvements in usability need not therefore produce corresponding gains in satisfaction \cite{hassenzahl_2018_ThingUnderstanding}. Although the nonsignificant result does not establish equivalent satisfaction across the two tools, it highlights the need to consider both the guidance provided by scaffolding and learners' need for autonomous control.

The absence of a significant difference in post-use affective experience similarly suggests that changing information presentation may not be sufficient to produce detectable affective benefits. Algorithm learning can involve difficulty, confusion, and frustration, and Pekrun et al.'s work on academic emotions emphasizes the roles of cognitive appraisals and motivational factors in affective experience \cite{pekrun_2002_AcademicEmotions}. The present comparison does not establish that either system is incapable of improving learners' emotions; it indicates only that no significant difference was observed. Research on collaborative approaches such as pair programming, in relation to anxiety and frustration in programming education \cite{salleh_2011_EmpiricalStudies}, suggests that affective support may also involve social interaction and the organization of learning beyond information presentation. These possibilities warrant further investigation.

Positive correlations among the five experience dimensions offer a complementary perspective. Usability was closely associated with perceived effectiveness, efficiency, and satisfaction, with the strongest association occurring between efficiency and satisfaction. Learners' evaluations thus comprise several interconnected dimensions. The association between perceived effectiveness and efficiency can be considered alongside Finneran's discussion of goal attainment and interaction experience \cite{finneran_2003_PersonArtefact}, while Tractinsky's work on aesthetics and usability judgments cautions against reducing overall experience to a single measure \cite{tractinsky_2000_WhatBeautiful}. These associations also resonate with the relationship between ease of use and usefulness in the Technology Acceptance Model \cite{venkatesh_2000_TheoreticalExtension}. Correlations do not, however, establish that improving one dimension will produce cascading improvements in others, nor do they explain the causal mechanisms underlying differences between systems. Taken together, the findings suggest that structured scaffolding supports clearer interactions that learners perceive as more helpful, while its benefits for efficiency, satisfaction, and affect require further examination.

\subsection{Experience and Affective Vulnerability}

Beyond the overall means, the subgroup findings suggest that the benefits of structured scaffolding should be interpreted in relation to learners' prior experience. In the standard AI condition, participants who rarely used chatbots gave lower User Adoption and Learning Management ratings, whereas no significant frequency-related differences were reported for VCC. This pattern is consistent with the possibility that structured support lowers experience-related barriers, but it does not establish that VCC eliminated differences associated with prior experience.

Open-ended chatbot interaction typically requires users to formulate questions, guide the conversation, and assess whether responses meet their learning needs. Prior use may therefore be associated with smoother adoption and learning management. Research on interface experience and interaction anxiety provides context for this interpretation \cite{rapp_2021_HumanSide}. Through predefined steps, algorithm walkthroughs, and synchronized visualizations, VCC takes on some of the work of organizing learning and may reduce the burden of constructing an interaction path independently. This interpretation aligns with the educational role of scaffolding in bridging knowledge gaps \cite{puntambekar_2005_ToolsScaffolding}. Nevertheless, prompting skills were not directly measured, and use frequency should not be equated with expertise. Prior research also indicates that the interaction limitations of conversational agents can frustrate even experienced users \cite{yang_2019_UnderstandingAffective}, so familiarity alone should not be treated as sufficient for a positive experience.

Unlike the pattern for use frequency, prior affective experience was associated with several evaluation dimensions in both systems. Here, affective background refers to participants' previous comfortable or uncomfortable experiences with chatbots and is distinct from the Emotion scale administered after the present interaction. In the standard AI condition, significant differences generally occurred between participants reporting neutral and comfortable prior experiences. In VCC, lower evaluations were more often concentrated among participants reporting uncomfortable prior experiences. These patterns suggest that reducing experience-related entry barriers does not necessarily ensure comparable interaction experiences across learners with different affective backgrounds.

This pattern was evident in interaction and layout evaluations. In standard AI, participants with neutral prior experiences rated interaction and layout lower than those with comfortable experiences. In VCC, participants with uncomfortable prior experiences rated interaction and layout lower than those with neutral or comfortable experiences, and rated the interface lower than those with comfortable experiences. Research on affect and attentional resources \cite{gable_2010_BluesBroaden}, together with processing fluency theory \cite{reber_2004_ProcessingFluency}, offers a possible interpretive lens: learners' existing affective appraisals and interaction expectations may be associated with how they understand and accept a new interface. The study did not measure momentary changes in attention, disorientation, or affective feedback loops, however, so these mechanisms remain possible explanations rather than established findings.

Information completeness showed a related pattern. In standard AI, participants with neutral prior experiences gave lower ratings than those with comfortable experiences; in VCC, those with uncomfortable prior experiences rated completeness lower than those with neutral experiences. Research on affect and information seeking suggests that affective appraisals may be related to how users judge whether information meets their goals \cite{kort_2001_AffectiveModel}. Structured presentation may also support perceived completeness by helping users locate and organize content \cite{mayer_2002_MultimediaLearning}. Yet the remaining differences associated with uncomfortable prior experiences in VCC indicate that clear content organization cannot be equated with all learners perceiving adequate support for learning.

The response consistency and efficiency findings further indicate that this pattern did not extend to every measure. In standard AI, participants with neutral prior experiences rated response consistency lower than those with comfortable experiences, whereas no corresponding significant comparison was reported for VCC. The Affect Infusion Model offers a theoretical perspective on the relationship between affective background and evaluative judgments \cite{forgas_1995_MoodJudgment}, but the present findings do not support attributing consistency differences to extreme emotions or cognitive changes under stress. VCC's predefined sequence may make interaction more predictable, although the absence of significant differences does not establish equivalent response reliability across affective backgrounds. The relationship between predictability and interaction naturalness remains a topic for further investigation \cite{luger_2016_HavingReally}.

Response time evaluations were associated with prior affective experience in both systems, although the significant comparisons involved different subgroups. Significant differences in information retrieval were reported only for standard AI, where participants with uncomfortable or neutral prior experiences gave lower ratings than those with comfortable experiences. VCC's sequential presentation may reduce the burden of formulating queries, consistent with cognitive load explanations of how structured guidance supports novices \cite{kalyuga_1998_LevelsExpertise}. At the same time, information retrieval research suggests that task complexity may also shape user experience \cite{lopatovska_2011_TheoriesMethods}. The present results are therefore best understood as associations between subjective efficiency evaluations and affective background. They do not establish equal objective response times, improved retrieval performance, or significantly weaker dependence on affect in either system.

The Learning Management findings extend this pattern to satisfaction. Standard AI showed evaluation differences associated with both use frequency and prior affective experience. For VCC, significant comparisons were reported only for prior affective experience, with participants reporting uncomfortable experiences giving lower ratings than those reporting comfortable experiences. Considered alongside research on learning interaction and affective engagement \cite{dmello_2012_AutoTutorAffective}, one possible interpretation is that predefined sequences can take on some of the work of organizing learning without necessarily resolving learners' existing discomfort or negative expectations about AI interaction. This distinction underscores the value of examining support for initial adoption separately from the experience of managing learning over time.

Together, these findings provide exploratory support for the proposed equity–vulnerability tension: structured design may lower interaction barriers associated with prior use while remaining subject to evaluation differences associated with uncomfortable prior experiences. Equity here is limited to opportunities to access interactional support across experience backgrounds; it does not imply that broader educational equity has been achieved. Affective vulnerability is likewise an interpretive framework, rather than a conclusion that structured design necessarily causes emotional harm. In particular, a significant difference within one system and a nonsignificant difference within another do not themselves constitute statistical evidence that the systems differ in their effects. Establishing this tension requires direct tests of interactions between system condition and learner background, alongside longitudinal or experimental designs that distinguish prior affective experience from emotional changes during use.

\subsection{Contextual Limits of Inclusive Design}

The value of structured visual support should be understood in relation to the devices through which it is accessed. In VCC, participants using laptops rated the interface more highly than those using tablets or smartphones; no significant device-related comparison was reported for standard AI. This finding suggests that the experience of a visual learning interface may depend on its presentation and interaction conditions. It does not establish that device-related effects are exclusive to VCC or that the two systems differ in their sensitivity to device type.

Display space offers one possible explanation. Algorithm visualizations may require learners to consult procedural steps, state changes, and textual explanations together, and limited display space may increase the effort needed to locate information and switch between representations. Research on small screens and the comprehension of graphical information provides context for this account \cite{hsieh_2016_EffectScreen}. However, the study compared device types without independently manipulating or measuring the contribution of screen size. Input methods, layout adaptation, and the surrounding use environment may also have shaped interface evaluations. The finding therefore identifies device-related differences in experience rather than a causal effect of screen size.

This result adds a contextual boundary to the preceding discussion of experience and affective background. Even when a system offers clear scaffolding or shows fewer experience-related differences, its support is realized through particular devices and interaction conditions. Inclusive design therefore cannot be assessed solely through average usability or barriers to initial adoption; it also requires examining whether different learners can continue to access support across conditions of use. The present study identified evaluation differences associated with affective background and device type separately, but did not test their joint effects. It consequently does not establish that smaller screens amplify affective vulnerability.

From this perspective, we propose extending the focus from initial accessibility to sustained support, using resilience to describe a system's capacity to maintain support as learner states and conditions of use change. Resilience is a theoretical and evaluative direction motivated by these findings, rather than a system property measured in this study. Future work should combine comparisons across devices, observations of sustained use, and measures of affect during interaction to examine whether the benefits of structured scaffolding persist under changing conditions. Lowering entry barriers and sustaining interactional support thus emerge as related design objectives for educational AI. The following section develops their implications for design, teaching, and deployment.

\section{Recommendations}

VCC's higher ratings for usability and perceived effectiveness suggest a targeted benefit of structured visual support, while no significant overall differences were detected for efficiency, satisfaction, or post-use emotion. These averages, however, do not show whether learners with different prior experiences or devices can continue to use that support effectively. We therefore recommend treating clear entry, sustained access to support, and independently demonstrated understanding as connected design and deployment goals.

First, structured visualization should elicit reasoning. Prior work shows that the learning value of algorithm visualization depends strongly on how learners engage with it \cite{hundhausen_2002_MetaStudyAlgorithm,naps_2002_ExploringRole}. Designers could embed short prediction--observation--explanation cycles into existing activities. In a cycle-detection task, for example, learners could predict which node or edge will be visited next, observe the resulting state change, and explain any mismatch. This makes the visualization a resource for checking reasoning rather than replacing it. Because this study measured perceived rather than objective learning outcomes, this pattern should be tested as a candidate learning mechanism.

Second, learners should have clear, learner-controlled ways to recover when reasoning breaks down. Instead of forcing progress through a fixed sequence, the system could let learners re-examine the current step, request a local hint, or return to the previous graph state while preserving progress. This direction is consistent with the study's within-group associations involving prior chatbot use and prior affective experience with chatbots. Prior work similarly highlights the need to balance access to assistance with continued cognitive engagement \cite{kazemitabaar_2024_CodeAidEvaluating,koedinger_2007_ExploringAssistance}. Support should help learners resolve useful confusion rather than eliminate all difficulty \cite{dmello_2014_ConfusionCan}. The present data support learner-controlled assistance as a design direction, but not automatic emotion recognition or affect-based intervention.

Third, tool-supported learning should be paired with independent checks of conceptual understanding. Both study conditions included instructor demonstration, partial guidance, and later self-directed use, so these recommendations apply to organized classroom deployment. Educators can model reasoning-focused questions and how to request clarification, but perceived effectiveness should not be treated as conceptual mastery. After an AI-supported activity, instructors could ask learners to solve a new graph problem without AI and explain how DFS identifies a cycle and why the relevant states change. Related studies show why learner experience, task performance, and independent understanding should be evaluated separately \cite{bassner_2026_LessStress,bastani_2025_GenerativeAI}. Deployment decisions should therefore consider both learner experience and independent performance.

Fourth, institutions should evaluate whether support is usable in representative learning conditions, not merely whether access is provided. Course-level pilots should include learners with different prior chatbot experiences and devices, and should record whether key steps can be completed, whether learners can resume after requesting help, how they experience the interaction, and whether they can later demonstrate understanding without AI. This follows the distinction between access, use, and outcomes in educational technology \cite{warschauer_2010_NewTechnology}. If particular devices or layouts repeatedly obstruct core tasks, institutions should revise the interface and, meanwhile, provide devices or alternative paths that support the same learning objective. Any collection of interaction or affect-related data should also have a clear purpose and protection plan.

Finally, these practices should be treated as testable design and deployment hypotheses. Future studies should directly test interactions between system condition and prior chatbot use, prior affective experience, and device type, rather than infer differences from significance in one subgroup but not another \cite{gelman_2006_DifferenceSignificant}. They should also examine repeated use, recovery after interruption, in-task experience, and delayed or no-AI assessments. The practical aim is to preserve learners' ability to seek and use support across different experiences and devices while testing whether that support leads to understanding they can demonstrate independently.

\section{Conclusion}
In our study, VCC received higher ratings than standard AI for usability and perceived effectiveness, while no significant differences were detected in efficiency, satisfaction, or post-use emotion. Infrequent chatbot users rated User Adoption and Learning Management lower within the standard AI condition, whereas no corresponding significant frequency-related comparisons were reported for VCC. Within VCC, learners reporting uncomfortable prior chatbot experiences rated interaction and layout less favorably than other learners, and laptop users rated the interface more highly than tablet or smartphone users. Together, these exploratory patterns distinguish the possibility of easing experience-related entry barriers from the question of whether support remains usable across affective backgrounds and conditions of use. The proposed equity--vulnerability tension is therefore an evaluative lens, rather than a causal effect established by these data. Future evaluations should test condition-by-background interactions directly, examine experience across devices, and assess learning independently of self-reports. Ultimately, structured support should be judged by how reliably learners can use it across different circumstances and whether they can demonstrate understanding beyond the AI interaction.

\begin{acks}
We sincerely thank all participants for their time and contributions to this study. We are also grateful to the reviewers for their thoughtful feedback and helpful suggestions, which improved this work.
\end{acks}

\bibliographystyle{ACM-Reference-Format}
\bibliography{reference}



\end{document}